\documentstyle[twoside,amssymb,12pt]{article}

\newcommand{\nc}{\newcommand}
\nc{\la}{\lambda} \nc{\alf}{\alpha}
\nc{\tht}{\theta}  \nc{\be}{\beta}  \nc{\eps}{\epsilon} \nc{\ze}{\zeta}
\nc{\ga}{\gamma}  \nc{\De}{\Delta}  \nc{\Ga}{\Gamma}  \nc{\vphi}{\varphi}
\nc{\de}{\delta} \nc{\si}{\sigma}  \nc{\ka}{\kappa}   \nc{\Si}{\Sigma}
\nc{\om}{\omega}  \nc{\qq}{\quad\quad}                \nc{\Om}{\Omega}
\nc{\nf}{\infty}   \nc{\dl}{\mathop{\smash{\cal L}}}  \nc{\black}{\rule{3mm}{3mm}}
\nc{\ra}{\rightarrow}  \nc{\ol}{\overline}  \nc{\und}{\underline}
\nc{\beq}{\begin{equation}}  \nc{\pt}{\partial}  \nc{\nin}{\noindent}
\nc{\eeq}{\end{equation}}
\nc{\beqa}{\begin{eqnarray}}  \nc{\dst}{\displaystyle}
\nc{\eeqa}{\end{eqnarray}} \nc{\nnb}{\nonumber}
\nc{\bs}{\backslash}        \nc{\mb}{\mathbb}

\newcounter{muni}
\newenvironment{remunerate}{\begin{list}{{\rm \arabic{muni}.}}
{\usecounter{muni}
\setlength{\leftmargin}{0pt}\setlength{\itemindent}{38pt}}}{\end{list}}

\nc{\brm}{\begin{remunerate}}   \nc{\erm}{\end{remunerate}}

\newtheorem{nlem}{Lemma} \newtheorem{nth}{Theorem}
\nc{\stg}{\mathop{\smash{*}}}
\nc{\st}{\mathop{\smash{\delta}}}
\nc{\barr}{\begin{array}}   \nc{\earr}{\end{array}}   \nc{\dg}{\dagger}
\nc{\mtvb}{\mathversion{bold}}   \nc{\mtvn}{\mathversion{normal}}
\nc{\ti}{\tilde}  \nc{\wti}{\widetilde} \nc{\wh}{\widehat}

\begin{document}

\begin{titlepage}
\begin{flushright}
February 2010
\end{flushright} 
\vskip 2.0truecm
\centerline{\large \bf EXPLICIT INTEGRABLE SYSTEMS}
\vskip 0.3truecm
\centerline{\large \bf ON TWO DIMENSIONAL MANIFOLDS}
\vskip 0.3truecm
\centerline{\large \bf WITH A CUBIC FIRST INTEGRAL}
\vskip 1.0truecm 
\centerline{\bf Galliano Valent${}^{*}$}
\vskip 2.0truecm 

\centerline{${}^{*}$\it Laboratoire de Physique Th\'eorique et des
Hautes Energies}
\centerline{\it Unit\'e associ\'ee au CNRS UMR 7589}
\centerline{\it 2 Place Jussieu, 75251 Paris Cedex 05, France} 
\vskip 3truecm

\begin{abstract}
A few years ago Selivanova gave an existence proof for some integrable models, in fact 
geodesic flows on two dimensional manifolds, with a cubic first integral. However the 
explicit form of these models hinged on the solution of a nonlinear third order ordinary 
differential equation which could not be obtained. We show that an appropriate choice of 
coordinates allows for integration and gives the explicit local form for the full family 
of integrable systems. The relevant metrics are described by a finite number of parameters 
and lead to a large class of models on the manifolds ${\mb S}^2,\,{\mb H}^2$ 
and $P^2({\mb R})$ containing as special cases examples due to Goryachev, Chaplygin, Dullin, 
Matveev and Tsiganov. 
\end{abstract}
\end{titlepage}

\nc{\pth}{P_{\tht}}  \nc{\pf}{P_{\phi}}  \nc{\pr}{P_{\rho}} \nc{\px}{P_x} \nc{\pz}{P_{\ze}}

\section{Introduction}
Let $M$ be a $n$-dimensional smooth manifold with metric $g(X,Y)=g_{ij}\,X^i\,Y^j$ and let 
$T^*\,M$ be its cotangent bundle with coordinates $(x,P)$, where $P$ is a covector 
from $T^*_x\,M$. Let us recall that $T^*M$ is a smooth symplectic $2n$-manifold with respect 
to the standard 2-form $\om=dP_i\wedge dx^i$ which induces the Poisson bracket
\[\{f,g\}=\sum_{i=1}^n\left(\frac{\pt f}{\pt x^i}\,\frac{\pt g}{\pt P_i}-
\frac{\pt f}{\pt P_i}\,\frac{\pt g}{\pt x^i}\right).\]
In $T^*\,M$ the geodesic flow is defined by the Hamiltonian 
\beq\label{notH}
H=K+V,\qq K=\frac 12\sum_{i,j=1}^n\,g^{ij}(x)\,P_i\,P_j,\qq V=V(x),\eeq
where $g^{ij}$ is the inverse metric of $g_{ij}$.

An ``observable" $f:\ T^*M\to {\mb R}$, which can be written locally 
\[f=\sum_{i_1+\cdots+i_n\leq m}\,f^{i_1,\cdots,i_n}(x)\,P_{i_1}\,\cdots P_{i_n},\qq \#(f)=m,\]
is a constant of motion  iff $\{H,f\}=0$. A hamiltonian system is said to be integrable in 
Liouville sense if there exist $n$ constants of motion (including $H$) generically independent 
and in pairwise involution with respect to the Poisson bracket.

In what follows we will deal exclusively with integrable systems defined on two dimensional manifolds. An integrable system is just made out of 2 independent observables $H$ and $Q$ with $\{H,Q\}=0$.
 
The general line of attack of this problem is based on the integer $m=\#(Q)$. For $m=1$ $M$ is 
a surface of revolution and for $m=2$ $M$ is a Liouville surface \cite{Da}. 

For higher values 
of $m$ only particular examples have been obtained, some of which in explicit form. For  
$M={\mb S}^2$ and $m=3$ the oldest explicit examples (early twentieth century) were due to 
Goryachev and Chaplygin on the one hand and to Chaplygin on the other hand 
(see \cite{bkf}[p. 483] and \cite{Ts} for the detailed references). On the same manifold 
with $m=4$ there is the famous Kovalevskaya system \cite{Ko} and some extension 
due to Goryachev (see \cite{Se4} for the reference).

More recently there was a revival of this subject due to Selivanova \cite{Se}, \cite{Se4} 
and Kiyohara \cite{Ki} who proved {\em existence} theorems of integrable systems for 
$m=3,4$ for the first author and for any $m\geq 3$ for the second author. As observed by 
Kiyohara himself for $m=3$ the two classes of models are markedly different. Even more 
recently several new explicit examples for $m=3$ were given by Dullin and Matveev \cite{dm} 
and Tsiganov \cite{Ts}. 
 
In this work we will focus on Selivanova's integrable systems with a cubic first integral 
discussed in \cite{Se}. The existence theorems she proved are not explicit since there remains 
to solve a nonlinear ODE of third order. In Tsiganov's article too a non-linear ODE of 
fourth order four appears for which only special solutions could be obtained. 

We will show that the solutions of these ODE are not required: the use of appropriate 
coordinates allows to get locally the explicit form of the full family of integrable system. 
Having the local form of the metric $g$ on $M$ one can determine the global structure of the manifold. In view of the many parameters exhibited by the metric, the global analysis gives 
rise to plenty of integrable models, some of which were discovered only recently.

The plan of the article is the following: in Section 2 we consider the class of models 
analyzed by Selivanova with the following leading terms for the cubic observable:
\[Q=p\,\pf^3+2q\,K\,\pf+\cdots,\qq p\in{\mb R},\quad q\geq 0,\]
and the general differential system resulting of $\{H,Q\}=0$ is given.

In section 3 we first integrate the special case where $q=0$: the differential system is reduced to 
a second order non-linear ODE. Its integration gives the local form of the integrable system and 
the global analysis detrmines the manifolds according to the parameters that appeared in the integration process. 

In Section 4 we consider the general case $q>0$. Here we have linearized, by an appropriate choice 
of the coordinates, the possibly non-linear ODE of third order. In Section 5, with the explicit 
local form of the metric, it is then straightforward (but lengthy because an enumeration of 
cases is required) to determine on which manifolds the metric is defined. We check that all 
the previously explicitly known integrable examples are indeed recovered.

\section{Cubic first integral}
Let us consider the hamiltonian (\ref{notH})
with
\beq 
K=\frac 12\Big(\pth^2+a(\tht)\pf^2\Big),\qq V=f(\tht)\cos\phi+g(\tht),\qq f(\tht)\not\equiv 0,
\eeq
and the cubic observable
\beq
Q=Q_3+Q_1,\eeq
with
\beq\left\{\barr{l}
Q_3=p\,\pf^3+2q\,K\,\pf,\qq p\,\in{\mb R},\ q\geq 0,\\[4mm] Q_1=\chi(\tht)\sin\phi\,\pth+\Big(\be(\tht)+\ga(\tht)\cos\phi\Big)\pf.\earr\right.\eeq

\begin{nlem} The constraint $\{H,Q\}=0$ is equivalent to the following differential system:
\beq\label{eq0}\barr{lc}
(a)\quad & \chi\,\dot{f}=\ga f,\qq \chi \dot{g}=\be\,f,\qq\quad\Big(\ \dot{}=D_{\tht}\Big),\\[4mm] 
(b)\quad & \dst\dot{\chi}=-q\,f,\qq\dot{\be}=2q\,\dot{g},\qq \dot{\ga}+\chi\,a=2q\,\dot{f},\qq a\ga+\chi\,\frac{\dot{a}}{2}=3(p+qa)f.\earr\eeq
\end{nlem}

\nin{\bf Proof:} 
The relation $\{H,Q\}=0$ splits into three constraints
\beq
\{K,Q_3\}=0,\qq  \{K,Q_1\}+\{V,Q_3\}=0,\qq  \{V,Q_1\}=0.\eeq
The first is identically true, the second one is equivalent to the relations (\ref{eq0}\,b) while the last one is equivalent to (\ref{eq0}\,a) .$\quad\Box$

The special case $q=0$ is rather difficult to obtain as the limit of the general case $q\neq 0$, 
so we will first work it out completely.

\section{The special case $q=0$}
We can take $p=1$ and obvious integrations give 
\beq\label{sc1}
\chi=\chi_0>0,\qq \be=\be_0\in{\mb R},\qq \ga=\chi_0\,\frac{\dot{f}}{f},
\qq \dot{g}=\frac{\be_0}{\chi_0}\,f,\qq a=-\frac{\dot{\ga}}{\chi_0},
\eeq
and the last equation
\beq\label{sc2}
\ddot{\ga}+2\,\frac{\dot{f}}{f}\,\dot{\ga}+6f=0.\eeq
An appropriate choice of coordinates does simplify matters:
 
\begin{nlem} The differential equation for $u=\dot{f}$ as a function of the 
variable $x=f$ is given by
\beq\label{eqd}
u\left(\frac{uu'}{x}\right)'+cx=0,\qq c=\frac 6{\chi_0}>0.\qq\qq\qq \Big(\ '=D_x \Big).\eeq 
\end{nlem}

\nin{\bf Proof:} 
The relations in (\ref{sc1}) become
\beq
g'=\frac{\be_0}{\chi_0}\,\frac xu,\qq \ga=\chi_0\,\frac ux,\qq a=-u\left(\frac ux\right)',\eeq
and (\ref{sc2}) gives (\ref{eqd}).$\quad\Box$

The solution of this ODE follows from
\begin{nlem} The general solution of (\ref{eqd}) is given by
\beq\label{eqds1}
u=-\frac{\ze^2+c_0}{2c},\eeq
with
\beq\label{eqds2}
\ze^3+3c_0\,\ze-2\rho=0,\qq\qq 2(\rho-\rho_0)=3c^2x^2,
\eeq
and integration constants $(\rho_0,c_0)$. 
\end{nlem}

\nin{\bf Proof:} 
Let us define $\,\dst \ze'=-c\,x/u$. This allows a first integration of  (\ref{eqd}), giving
$\,\dst\frac{uu'}{x}=\ze$. From this we deduce
\[cu'=-\ze\ze'\qq\Longrightarrow\qq  2c\,u=-\ze^2-c_0,\]
which in turn implies
\beq
\Big[\ze^2+c_0\Big]\ze'=2c^2\,x\quad\Longrightarrow\quad 
\ze^3+3c_0\,\ze-2\rho=0,\eeq
which concludes the proof. $\quad\Box$

It is now clear that the initial coordinates $(\tht,\phi)$ chosen on $S^2$ will not lead, 
at least generically, to a simple form of the hamiltonian! To achieve a real simplification 
for the observables the symplectic coordinates change 
$(\tht,\phi,\pth,\pf)\ \to\ (\ze,\phi,\pz,\pf)$ gives: 

\begin{nth}\label{th1} 
Locally, the integrable system has for explicit form
\beq\label{sys0}
\left\{\barr{l}\dst 
H=\frac 12\left(F\,\pz^2+\frac{G}{4F}\,\pf^2\right)+\chi_0\,\sqrt{F}\,\cos\phi
-\be_0\,\ze,\\[5mm]\dst 
Q=\pf^3-2\chi_0\Big(\sqrt{F}\,\sin\phi\,\pz+(\sqrt{F}\,)'\,\cos\phi\,\pf\Big)
+2\be_0\,\pf,\earr\right.\qq \Big(\ '=D_{\ze}\Big),\eeq
with
\beq\label{FetG}
F=-2\rho_0+3c_0\,\ze+\ze^3,\qq\qq G=9c_0^2+24\rho_0\,\ze-18c_0\,\ze^2-3\ze^4.\eeq
\end{nth}

\nin{\bf Proofs:} 
One may obtain these formulas by elementary computations and some scalings of $\chi_0,\,\be_0$ 
and $H$.
 
Alternatively, one can check that (\ref{FetG}) implies the relations   
\beq\label{usid0}
G'=-12\,F,\qq G=F'^2-2F\,F'',\eeq
which allows for a direct check of $\,\{H,Q\}=0$.
As proved in \cite{Se} this system does not exhibit any linear or quadratic constant of motion 
and $\,(H,Q)$ are algebraically independent.
$\quad\Box$

We are now in position to analyze the global geometric aspects related to the metric
\beq\label{met}
g=\frac{d\ze^2}{F}+\frac{4F}{G}\,d\phi^2,\qq\quad\phi\in [0,2\pi).\eeq

One has first to impose the positivity of both $F$ and $G$ for this metric to be riemannian. 
This gives for $\ze$ some interval $I$ whose end-points are possible singularities of the metric. 
To ascertain that the metric is defined on some manifold one has to ensure that these 
singularities are apparent ones and not true curvature singularities. 

Let us define, for the cubic $F$, its discriminant $\De=c_0^3+\rho_0^2$. 

\begin{nth}\label{th2}
The metric (\ref{met}): 
\brm
\item[(i)] For $\De<0$ is defined on ${\mb S}^2$ iff
\[F=(\ze-\ze_0)(\ze-\ze_1)(\ze-\ze_2),\qq\quad \ze_0<\ze<\ze_1<\ze_2.\]
The change of coordinates
\beq\label{notq0} 
{\rm sn}\,(u,k^2)=\sqrt{\frac{\ze-\ze_0}{\ze_1-\ze_0}},\qq  
k^2=\frac{\ze_1-\ze_0}{\ze_2-\ze_0}\in\,(0,1),\eeq 
gives for integrable system \footnote{We use the shorthand notation: $s,\,c,\,d$ respectively 
for ${\rm sn}\,(u,k^2),\,{\rm cn}\,(u,k^2),\,{\rm dn}\,(u,k^2)$.}
\beq\label{sys1q0}\left\{\barr{l}\dst
H=\frac 12\left(P_u^2+\frac{D(u)}{s^2 c^2 d^2}\pf^2\right)+\chi_0\,k^2\,s c d\,\cos\phi
-\be_0\,k^2 s^2,\\[5mm]\dst 
Q=4\pf^3-\chi_0\left(\sin\phi\,P_u+\frac{(scd)'}{scd}\,\cos\phi\,\pf\right)+2\be_0\,\pf,\\[5mm] 
D(u)=(1-k^2 s^4)^2-4k^2\,s^4c^2d^2,\qq u\in(0,K).\earr\right.\eeq
\item[(ii)] For $\De=0$ is defined on ${\mb H}^2$ iff
\[F=(\ze-\ze_1)^2(\ze+2\ze_1),\qq\quad -2\ze_1<\ze<\ze_1,\quad (\ze_1>0).\]
The change of coordinates
\beq\label{notspeq0}
\ze=\ze_1(-2+3\,\tanh^2 u),\qq u\in(0,+\nf),\eeq
gives for integrable system \footnote{We use the shorthand notation $S,\,C,\,T$ respectively for $\sinh u,\,\cosh u,\,\tanh u$.}
\beq\label{sys2q0}
\left\{\barr{l}\dst
H=M_1^2+M_2^2-\left(1-\frac 3{C^2}\right)M_3^2+\chi_0\,T(1-T^2)\,\cos\phi-\be_0\,T^2,\\[5mm]\dst
Q=4M_3^3-\chi_0\Big(M_1-3T\,\cos\phi\,M_3\Big)+2\be_0\,M_3.\earr\right.\eeq
\item[(iii)] For $\De>0$ is not defined on a manifold.
\erm
\end{nth}

\nin{\bf Proof of (i):}
If $\,\De<0$ the cubic $F$ has three simple real roots $\ze_0<\ze_1<\ze_2$. 
If we take $\,\ze\in(\ze_2,+\nf)$ then $F$ is positive. The relation $\,G'=-12F$ shows 
that in this interval $G$ is decreasing from $G(\ze_2)=F'^2(\ze_2)>0$ to 
$-\nf$ and will vanish for some $\wh{\ze}>\ze_2$. Hence to ensure positivity for $F$ and $G$ 
we must restrict $\ze$ to the interval $(\ze_2,\wh{\ze})$. Since at $\ze=\wh{\ze}$ the 
function $F$ does not vanish while $G$ does, this point is a curvature 
singularity and the metric cannot be defined on a manifold.

The positivity of $F$ is also ensured if we take $\ze\in(\ze_0,\ze_1)$. In this 
interval $G$ decreases monotonously from $G(\ze_0)$ to $\,G(\ze_1)=F'^2(\ze_1)>0$. 
Let us analyze the singularities at the end points. For $\ze$ close to $\ze_0$ we have for 
approximate metric
\beq
g\approx \frac 4{F'(\ze_0)}\left[\frac{d\ze^2}{4(\ze-\ze_0)}
+\frac{F'^2(\ze_0)}{G(\ze_0)}\,(\ze-\ze_0)\,d\phi^2\right].
\eeq
The relation (\ref{usid0}) gives $G(\ze_0)=F'^2(\ze_0)$, so the change of variable 
$\rho=\sqrt{\ze-\ze_0}$ allows to write
\beq
g\approx \frac 4{F'(\ze_0)}\Big(d\rho^2+\rho^2\,d\phi^2\Big),\eeq
which shows that $\rho=0$ is an {\em apparent} singularity, due to the choice of polar coordinates, which could  removed by switching back to cartesian coordinates. So the point $\ze=\ze_0$ can be added to the manifold. 

A similar argument works for $\ze=\ze_1$. In fact these end-points are geometrically the poles of the  manifold and the index theorem for $\,\pt_{\phi}$ gives for Euler characteristic $\,\chi=2$,  showing that the manifold is indeed ${\mb S}^2$. 
Then, using the change of variable (\ref{notq0}), it is a routine exercise in 
elliptic functions theory to operate the symplectic coordinates change 
$\,(\ze,\phi,\pz,\pf)\ \to\ (u,\phi,P_u,\pf)$ which, after several scalings of the 
observables and of their parameters, gives (\ref{sys1q0}). Notice that one can also, by direct computation, check that $\,\{H,Q\}=0$ from the formulas given in (\ref{sys1q0}). $\quad\Box$

\vspace{2mm}
\nin{\bf Proof of (ii):}
In this case we have
\[F=(\ze+2\ze_1)(\ze-\ze_1)^2,\qq\quad G=-3(\ze+3\ze_1)(\ze-\ze_1)^3,\qq\qq \ze_1=-\rho_0^{1/3}.\]
For $\ze_1<0$ the positivity of $F$ implies $\ze\in(2|\ze_1|,+\nf)$ and $G$ decreases and vanishes for $\wh{\ze}=3|\ze_1|$ leading to a curvature singularity. The case 
$\ze_1=0$ is also excluded since then $G\leq 0$ and the remaining case is $\ze_1>0$. The 
positivity of $F$ and $G$ requires $\ze\in(-2\ze_1,\ze_1)$. The singularity structure is most 
conveniently discussed thanks to the coordinates change (\ref{notspeq0})  
which brings the metric to the form
\beq\label{methyp}
g=\frac 4{3\ze_1}\left\{du^2+\frac{\sinh^2 u}{1+3\tanh^2 u}\,d\phi^2\right\},\qq
u\in(0,+\nf),\eeq
from which we conclude that the manifold is ${\mb H}^2$.
Then starting from (\ref{sys0}), the symplectic change of coordinates $(\ze,\phi,\pz,\pf)\ \to\  (u,\phi,P_u,\pf)$, and some scalings, gives 
\beq\label{sys2bq0}\left\{\barr{l}\dst 
H=\frac 12\left(P_u^2+\frac{(1+3\,T^2)}{S^2}\,\pf^2\right)
+\chi_0\,T(1-T^2)\cos\phi-\be_0\,T^2,\\[5mm]\dst 
Q=4\pf^3-\chi_0\left(\sin\phi\,P_u+\frac{1-3T^2}{T}\,\cos\phi\,\pf
\right)+2\be_0\,\pf,\earr\right.\eeq
Defining the generators of the $so(2,1)$ Lie algebra in $T{\mb H}^2$ to be
\beq\label{genM}
M_1=\sin\phi\,P_u+\frac{\cos\phi}{T}\,\pf,\qq 
M_2=\cos\phi\,P_u-\frac{\sin\phi}{T}\,\pf,\qq M_3=\pf,\eeq
transforms the observables (\ref{sys2bq0}) into (\ref{sys2q0}).$\quad\Box$

\vspace{2mm}
\nin{\bf Proof of (iii):} 
For $\,\De>0$ the cubic $F$ has a single real zero $\ze_0$. The 
positivity of $F$ requires that $\ze\in\,(\ze_0,+\nf)$. Since $G'=-12F$ the function 
$G$ decreases from $G(\ze_0)$ to $-\nf$. Since $G(\ze_0)>0$ there exists $\wh{\ze}>\ze_0$ for 
which $G(\wh{\ze})=0$. So positivity restricts $\ze\in(\ze_0,\wh{\ze})$ and 
$\,\wh{\ze}$ is a curvature singularity showing that the metric cannot be defined 
on a manifold.$\quad\Box$

\nin{\bf Remarks:}
\brm
\item The integrable system (\ref{sys2q0}) corresponds to the limit of (\ref{sys1q0}) when 
$\ze_2\to\ze_1$ or $k^2\to 1$. Then the elliptic functions degenerate into hyperbolic functions. 
Let us emphasis that in this limit the observables behave smoothly while the manifold changes 
drastically . Let us also observe that $H$ is globally defined on the manifold while $Q$ is not.
\item In \cite{Se} Selivanova proved an existence theorem for an integrable system 
on $S^2$ with a cubic observable (case (i) of her Theorem 1.1). The observables are
\beq\label{eqSe1}\barr{l}\dst 
H=\frac{\psi'^2(y)}{2}\Big(P_y^2+\pf^2\Big)
+\frac{\psi'^2(y)}{2}(\psi(y)-\psi''(y))\,\cos\phi,\\[5mm]\dst
Q=\pf^3-\frac 32\,\psi'(y)\,\sin\phi\,P_y+\frac 32\,\psi(y)\,\cos\phi\,\pf,\earr
\qq\Big(\ '=D_y\Big),\eeq
where $\psi(y)$ is a solution of the ODE
\beq\label{eqSe2}
\psi'\,\psi''=\psi\,\psi''-2\psi''^2+\psi'^2+\psi^2,\qq \psi(0)=0,\ \psi'(0)=1,\ \psi''(0)=\tau.\eeq
Comparing (\ref{eqSe1}) and (\ref{sys0}) for $\be_0=0$ makes it obvious that 
we are dealing with the same integrable system, up to diffeomorphism. The 
{\em local} identification follows from  
\beq\label{Seli}
\psi(y)=-\frac{(\ze^2+c_0)}{2\sqrt{F}},
\qq\quad \frac{\sqrt{G}}{F}\,d\ze=\pm\sqrt{3}\,dy, \eeq
and we have checked that the ODE (\ref{eqSe2}) is a consequence of the 
relations (\ref{Seli}) and (\ref{usid0}). We see clearly that Selivanova's choice of the 
coordinate $y$ led to a complicated ODE, very difficult to solve. In fact one should rather 
find coordinates such that the ODE becomes tractable, as we did.
\erm

\section{Local structure of the integrable systems for $q>0$}
As already observed, if one insists in working with the variable $\tht$, the differential system (\ref{eq0}) can be reduced  either to a third order \cite{Se} or to a fourth order \cite{Ts} non-linear ODE. The key to a full integration of this system is again an appropriate choice of coordinates on the manifold. 

\begin{nth} 
Locally, the integrable system $\,(H,Q)$ has for explicit form
\beq\label{sysq}
\left\{\barr{l}\dst 
H=\frac 1{2\ze}\Big(F\,P_{\ze}^2+\frac G{4\,F}\,\pf^2\Big)
+\frac{\sqrt{F}}{2q\ze}\,\cos\phi+\frac{\be_0}{2q\ze},\\[5mm]\dst   
Q=p\,\pf^3+2q\,H\,\pf-\sqrt{F}\,\sin\phi\,P_{\ze}
-(\sqrt{F}\,)'\,\cos\phi\,\pf,\earr\right.\qq \Big(\ '=D_{\ze}\Big),\eeq
with the polynomials
\beq 
F=c_0+c_1\ze+c_2\ze^2+\frac pq\,\ze^3,\qq G=F'^2-2F\,F''.\eeq
\end{nth}

{\bf Proofs:}
Starting from (\ref{eq0}) the functions $\be$ and $g$ are easily determined to be
\beq 
\be=\frac{\be_0}{\chi^2},\qq g=\frac{\be_0}{2q\chi^2}.\eeq
The functions $\ga$ and $a$ can be expressed in terms of $f$ and its derivatives with 
respect to $\,\chi$ as
\beq
\ga=-q\chi\,f',\qq a=-q^2\left(ff''+\frac 3{\chi}\,ff'\right).\eeq
Then the last relation in (\ref{eq0}) reduces to a second order {\em linear} ODE
\beq
\chi\,(ff')''+9\,(ff')'+\frac{15}{\chi}\,ff'=\frac{6p}{q^3},\eeq
which is readily integrated to 
\beq\label{ff1}
f=\pm\sqrt{c_2+f_1\,\chi^2+\frac{c_1}{\chi^2}+\frac{c_0}{\chi^4}},\qq\qq f_1=\frac p{4q^3}.\eeq
The remaining functions become
\beq\label{aga}\barr{l}\dst 
a=\frac{q^2}{f^2}\left(\frac{c_1^2-4c_0c_2}{\chi^6}-\frac{12c_0f_1}{\chi^4}-\frac{6c_1f_1}{\chi^2}
-4c_2f_1-3f_1^2\,\chi^2\right),\\[5mm]\dst 
\ga=\frac q{f}\left(-f_1 \chi^2+\frac{c_1}{\chi^2}+\frac{2c_0}{\chi^4}\right).\earr\eeq
The observables can be written, up to a scaling of the parameters, in terms of $F$ and $G$ defined by
\beq\label{basdef}\barr{l}\dst 
F=4q^2\,\chi^4\,f^2=c_0+c_1\ze+c_2\ze^2+g_1\ze^3,\qq g_1=\frac pq,\qq \ze=\chi^2,\\[4mm]\dst 
G=16q^2\,\chi^6\,f^2\,a=c_1^2-4c_0c_2-12c_0g_1\ze-6c_1g_1\ze^2-4c_2g_1\ze^3-3g_1^2\ze^4.
\earr\eeq
To simplify matters the symplectic change of coordinates 
$(\tht,\phi,\pth,\pf)\ \to\ (\ze,\phi,P_{\ze},\pf)$.  
gives the required result, up to scalings.

Alternatively (\ref{basdef}) implies the relations     
\beq\label{usidq}
G'=-12\frac pq\,F,\qq G=F'^2-2F\,F'',\eeq
which allow a direct check of $\,\{H,Q\}=0$. As proved in \cite{Se} this system does not exhibit any 
other conserved observable linear or quadratic in the momenta, and $\,(H,Q)$ are 
algebraically independent.$\quad\Box$

\vspace{4mm}
\nin {\bf Remarks:}
\brm
\item The limit $q=0$ is quite tricky: it is why we analyzed it separately in the 
previous section.
\item Let us observe that the kinetic parts of $H$ in (\ref{sys0}) and (\ref{sysq}) are 
conformally related. 
\item It is still possible to come back to the coordinate $\tht$ but the price to pay is the 
integration of the relation
\beq
\sqrt{\frac{\ze}{F}}\,d\ze=-d\tht,\eeq
which can be done using elementary functions for $c_0=0$.
\erm

\section{The global structure}
Let us now examine the global geometric aspects of the metric
\beq\label{metq}
g=\frac{\ze}{F}\,d\ze^2+\frac{4\,\ze\,F}{G}\,d\phi^2,\qq\quad \phi\in[0,2\pi),\eeq
taking into account the following observations:
\brm
\item The positivity constraints are $\ze F(\ze)>0$ and $G(\ze)>0$. They define the 
end-points of some interval $I$ for $\ze$. In some cases, discussed in detail later on, 
one can obtain extensions beyond some of the end-points.
\item For the observables to be defined it is required that $\,F\geq 0\quad\forall\ze\in I$.
\item As already observed any point $\ze_0$ with $F(\ze_0)\neq 0$ and $G(\ze_0)=0$ is a curvature singularity.
\item The point $\ze=0$ is a curvature singularity for $F(0)\neq 0$ and $G(0)\neq 0$. 
\erm

In order to have a complete description of all the possible integrable models, we will present them 
in three sets:
\brm
\item The first set $p=0$ with a simpler geometric structure.
\item The second set $p>0$ somewhat similar to the $q=0$ case.
\item The third set $p<0$ which displays the richest structure. 
\erm 

\subsection{First set of integrable models}
Since $\,p=0$ we have
\beq\label{notp0}
F=c_0+c_1\ze+c_2\ze^2=c_2(\ze-\ze_1)(\ze-\ze_2),\qq G=c_1^2-4c_0c_2,
\qq\quad (c_0,\,c_1,\,c_2)\in{\mb R}^3.\eeq

\begin{nth} 
In this set we have the following integrable models:
\brm
\item[(i)] Iff $c_2>0$ and $0<\ze_2<\ze$ the metric (\ref{metq}) is defined in $\,{\mb H}^2$ and 
\beq\label{si1p0}\left\{\barr{l}\dst
H=\frac 12\,\frac{M_1^2+M_2^2-M_3^2}{\rho+\cosh u}
+\frac{\alf\,\sinh u\,\cos\phi+\be}{\rho+\cosh u},\qq u\in(0,+\nf),\\[5mm]\dst
Q=H\,M_3-\alf\,M_1,\qq \rho=\frac{\ze_2+\ze_1}{\ze_2-\ze_1}\in(-1,+\nf).\earr\right.\eeq
\item[(ii)] Iff $c_2<0$ and $\,0<\ze_1<\ze<\ze_2$ the metric 
(\ref{metq}) is defined in $\,{\mb S}^2$ and 
\beq\label{si2p0}\left\{\barr{l}\dst
H=\frac 12\,\frac{L_1^2+L_2^2+L_3^2}{1+\rho\cos\tht}
+\frac{\alf\,\rho\,\sin\tht\,\cos\phi+\be}{1+\rho\cos\tht},\qq\tht\in(0,\pi),\\[5mm]\dst
Q=H\,L_3+\alf\,L_1,\qq \rho=\frac{\ze_2-\ze_1}{\ze_2+\ze_1}\in(0,+1).\earr\right.\eeq
\item[(iii)] Iff $c_2=0$ the metric 
(\ref{metq}) is defined in $\,{\mb R}^2$ and 
\beq\label{si3p0}\left\{\barr{l}\dst
H=\frac 12\,\frac{P_x^2+P_y^2}{1+\rho^2(x^2+y^2)}
+\frac{2\alf\,\rho^2\,x+\beta}{1+\rho^2(x^2+y^2)},\qq (x,y)\in{\mb R}^2,\\[5mm]\dst
Q=H\,L_z-\alf\,P_y,\qq \rho>0.\earr\right.\eeq
\erm
In all cases $\alf$ and $\be$ are free parameters.
\end{nth}

\nin{\bf Proof of (i):}
The positivity condition $G>0$ shows that $F$ has two real and distinct roots $\ze_1<\ze_2$,  
so we will write
\beq
F=c_2(\ze-\ze_1)(\ze-\ze_2),\qq G=c_2^2(\ze_1-\ze_2)^2.\eeq
Then imposing the positivity of $\ze F$ one has to deal with the iff part of the proof by an enumeration 
of all possible cases for the triplet $(0,\ze_1,\ze_2)$, including the possibility of one $\ze_i$ 
being zero. Taking into account the remarks given at the end of Section 4, one concludes that for $c_2>0$, we must take $\ze>\ze_2>0$. The change of coordinates
\[\ze=\frac{\ze_2-\ze_1}{2}\Big(\rho+\cosh u\Big),\qq(\ze_2,+\nf)\ \to\ (0,+\nf),
\qq \rho=\frac{\ze_2+\ze_1}{\ze_2-\ze_1}.\]
brings the metric (\ref{metq}) to the form
\beq
g=\frac{\ze_2-\ze_1}{2c_2}\Big(\rho+\cosh u\Big)
\Big(du^2+\sinh^2 u\,d\phi^2\Big),\qq u\in(0,+\nf),\eeq
which is conformal to the canonical metric on ${\mb H}^2$. Using the definitions (\ref{genM}) 
we obtain (\ref{si1p0}), up to scalings.$\quad\Box$

\vspace{2mm}
\nin{\bf Proof of (ii):}
For $c_2<0$ positivity requires either $0<\ze_1<\ze<\ze_2$ or $\ze_1<\ze<\ze_2<0$. 
In both cases the change of coordinates
\[\ze=\frac{\ze_1+\ze_2}{2}\Big(1+\rho\,\cos\tht\Big),\qq (\ze_1,\ze_2)\to(\pi,0),
\qq \rho=\frac{\ze_2-\ze_1}{\ze_2+\ze_1},\]
brings the metric (\ref{metq}) to one and the same form
\beq
g=\frac{\ze_1+\ze_2}{2c_2}\Big(1+\rho\,\cos\tht\Big)
\Big(d\tht^2+\sin^2\tht\,d\phi^2\Big),\qq \tht\in(0,\pi),\eeq
which is conformal to the canonical metric on ${\mb S}^2$ for $\rho\in(0,+1)$. Using the $so(3)$ 
Lie algebra generators acting in $T^*{\mb S}^2$
\beq\label{genL}
L_1=\sin\phi\,P_{\tht}+\frac{\cos\phi}{\tan\tht}\,\pf,\qq 
L_2=\cos\phi\,P_{\tht}-\frac{\sin\phi}{\tan\tht}\,\pf,\qq L_3=\pf,\eeq
one obtains (\ref{si2p0}), up to scalings.$\quad\Box$

\vspace{2mm}
\nin{\bf Proof of (iii):}
For $c_2=0$ we have $G=c_1^2>0$. 

If $c_1<0$ we can write $F=|c_1|(\ze_1-\ze)$ and positivity 
requires $\ze\in(0,\ze_1)$. If $\ze_1\neq 0$ then $\ze=0$ is a curvature singularity because 
$F(0)$ and $G(0)$ are not vanishing.

If $c_1>0$ we have $F=c_1(\ze-\ze_1)$. If $\ze_1<0$ positivity requires either $\ze>0$, but $\ze=0$ 
is a curvature singularity, or $\ze<\ze_1$ and then $F$ is negative. If 
$\ze_1=0$ the metric becomes
\[g=\frac 1{c_1}\Big(d\ze^2+4\ze^2\,d\phi^2\Big),\]
so to recover flat space we have to take $\wti{\phi}=2\phi\,\in\,[0,2\pi)$ and in $H$ appears a term 
of the form $\cos(\wti{\phi}/2)$ which does not define a function in ${\mb R}^2$.
Eventually, if $\ze_1>0$ if we take $\ze<0$ the point $\ze=0$ is singular, so we are left with 
$\ze>\ze_1$. The change of coordinates
\[\ze=\ze_1(1+\rho^2\,r^2),\quad \rho>0,\qq x=r\cos\phi,\quad y=r\sin\phi,\]
brings the metric (\ref{metq}) to the form
\beq
g=\frac{4\ze_1^2\rho^2}{c_1}(1+\rho^2\,r^2)(dx^2+dy^2),\qq (x,\,y)\in{\mb R}^2.\eeq
Using the $e(3)$ Lie algebra generators $(P_x,P_y,L_z=xP_y-yP_x)$  we obtain (\ref{si3p0}), 
up to scalings.
$\quad\Box$

The remaining cases are given by $p\neq 0$. It is convenient to rescale $F$ 
by $|p|/q$ and $G$ by $p^2/q^2$ in order to have
\beq\label{newF}
F=\eps(\ze^3+f_0\ze^2+c_1\ze+c_0),\quad \eps={\rm sign}(p),\qq G=F'^2-2FF'',
\quad G'=-12\,\eps\,F,\eeq
and for the observables, up to scalings
\beq\label{newH}
\left\{\barr{l}\dst 
H=\frac 1{2\ze}\Big(F\,P_{\ze}^2+\frac G{4\,F}\,\pf^2\Big)
+\alf\,\frac{\sqrt{F}}{\ze}\,\cos\phi+\frac{\be}{\ze},\\[5mm]\dst   
Q=\eps\,\pf^3+2\,H\,\pf-2\alf\Big(\sqrt{F}\,\sin\phi\,P_{\ze}
+(\sqrt{F}\,)'\,\cos\phi\,\pf\Big),\earr\right.\eeq
So the metric is still given by (\ref{metq}). We will denote 
by $\De_{\eps}$ the discriminant of $F$ according to the sign of $\eps$.

\subsection{Second set of integrable models}
It is given by $\,p>0$ or $\,\eps=+1$. We have:

\begin{nth} 
The metric (\ref{metq}):
\brm
\item[(i)] For $\De_+<0$ is defined on ${\mb S}^2$ iff
\[F=(\ze-\ze_0)(\ze-\ze_1)(\ze-\ze_2),\qq\quad 0<\ze_0<\ze<\ze_1<\ze_2.\]
The integrable system, using the notations of Theorem 2 case (i), is
\beq\label{sys1q}\left\{\barr{l}\dst
H=\frac 1{2\ze_+(u)}\left(P_u^2+\frac{D(u)}{s^2 c^2 d^2}\pf^2\right)
+\alf k^2\frac{\,scd}{\ze_+(u)}\,\cos\phi+\frac{\be}{\ze_+(u)},\\[5mm]\dst 
Q=4\pf^3+2H\,\pf-\alf\left(\sin\phi\,P_u+\frac{(scd)'}{scd}\,\cos\phi\,\pf\right),\\[5mm] 
\ze_+(u)=\rho+k^2\,{\rm sn}^2\,u,\qq u\in\,(0,K), 
\quad \rho=\frac{\ze_0}{\ze_2-\ze_0}>0.\earr\right.\eeq 
\item[(ii)] For $\De_+=0$ is defined on ${\mb H}^2$ iff 
\[F=(\ze-\ze_0)(\ze-\ze_1)^2,\qq\quad 0<\ze_0<\ze<\ze_1.\]
The integrable system, using the notations of Theorem 2 case (ii), is
\beq\label{sys2q}
\left\{\barr{l}\dst
H=\frac 1{2\ze_+(u)}\left\{M_1^2+M_2^2-\left(1-\frac 3{C^2}\right)M_3^2\right\}+\alf\,\frac{T(1-T^2)}{\ze_+(u)}\,\cos\phi
+\frac{\be}{\ze_+(u)},\\[5mm]\dst
Q=4M_3^3+2H\,M_3-\alf\Big(M_1-3T\,\cos\phi\,M_3\Big),\\[5mm]
\ze_+(u)=\rho+\tanh^2 u,\quad u\in(0,+\nf),\qq \rho=\frac{\ze_0}{\ze_1-\ze_0}>0.
\earr\right.\eeq
\item[(iii)] For $\De_+>0$ is not defined on a manifold.
\erm\end{nth}

\nin{\bf Proof of (i):} The iff part results from a case by case examination of all possible 
orderings of the 4-plet $(0,\ze_0,\ze_1,\ze_2)$, including the possibility of one  
of the $\,\ze_i$ being zero. We will not give the full details which can be easily 
checked by the reader, taking into account the remarks presented at the end of Section 4. 
The reader can check that with  $F=(\ze-\ze_0)(\ze-\ze_1)(\ze-\ze_2)$ and 
$0<\ze_0<\ze<\ze_1<\ze_2$, the polynomial $F$ is positive and vanishes at the end-points 
$(\ze_0,\,\ze_1)$ while $G$ is strictly positive. It follows that $\ze=\ze_0$ and $\ze=\ze_1$ 
are poles and the manifold is ${\mb S}^2$. Operating the same coordinate change as 
in Theorem 2, case (i), one obtains (\ref{sys1q}).$\quad\Box$

\vspace{2mm}
\nin{\bf Proof of (ii):} 
The polynomial $G$ becomes $G=(\ze_1-\ze)^3(3\ze+\ze_1-4\ze_0)$. The change of variable
\[\ze=(\ze_1-\ze_0)(\rho+{\rm th}^2 u),
\qq (\ze_0,\ze_1)\to(0,+\nf),\qq\rho=\frac{\ze_0}{\ze_1-\ze_0}>0,\]
transforms the observables, up to scalings, into
\beq\left\{\barr{l}\dst 
H=\frac 1{2\ze_+(u)}\left(P_u^2+\frac{1+3T^2}{S^2}\,\pf^2\right)
+\frac{\alf}{\ze_+(u)}\,T(1-T^2)\,\cos\phi+\frac{\be}{\ze_+(u)},\\[5mm]\dst 
Q=4\pf^3+2H\,\pf-\alf\,\sin\phi\,P_u-\alf\,\frac{(1-3T^2)}{T}\,\cos\phi\,\pf,\\[5mm]
\ze_+(u)=\rho+\tanh^2 u.\earr\right.\eeq
Using the relations (\ref{genM}) one gets (\ref{sys2q}). $\quad\Box$

\vspace{2mm}
\nin{\bf Proof of (iii):}
Examining all the possible cases gives no manifold for the metric.$\quad\Box$

\subsection{Third set of integrable models}
It is given by $\,p<0$ or $\,\eps=-1$. It displays a richer structure and for 
clarity we will split up the description of the integrable systems into several theorems.

\begin{nth} 
The metric (\ref{metq}) for $\De_-<0$ is defined on ${\mb S}^2$ iff:
\brm
\item[(i)] either $\  F=(\ze-\ze_0)(\ze-\ze_1)(\ze_2-\ze),\quad  
\ze_0<\ze_1<\ze<\ze_2\ (\ze_1>0).$

\nin The change of coordinates
\beq\label{notq} 
{\rm sn}\,(u,k^2)=\sqrt{\frac{\ze_2-\ze}{\ze_2-\ze_1}}, \qq 
k^2=\frac{\ze_2-\ze_1}{\ze_2-\ze_0}\in\,(0,1),\eeq 
gives for integrable system 
\beq\label{sys1qm}\left\{\barr{l}\dst 
H=\frac 1{2\ze_-(u)}\left(P_u^2+\frac{D(u)}{s^2 c^2 d^2}\,\pf^2\right)
+\alf\frac{k^2\,scd}{\ze_-(u)}\,\cos\phi+\frac{\be}{\ze_-(u)},\\[5mm]\dst 
Q=-4\,\pf^3+2H\,\pf+\alf\left(\sin\phi\,P_u+\frac{(scd)'}{scd}\,\cos\phi\,\pf\right),\\[5mm]
\ze_-(u)=k^2\Big(\rho-{\rm sn}^2\,u\Big),\qq u\in(0,K),\qq\rho=\frac{\ze_2}{\ze_2-\ze_1}>1.
\earr\right.\eeq
\item[(ii)] or $\ F=(\ze_0-\ze)(\ze-\ze_1)(\ze-\ze_2)\ \mbox{and}\ \ G(0)=0,
\quad 0<\ze<\ze_0<\ze_1<\ze_2.$

\nin The integrable system is 
\beq\label{sys2qm}
\left\{\barr{l}\dst 
H=\frac 12\,f\,(L_1^2+L_2^2)+\frac 12\left(\frac{h}{3f}-\cos^2\tht\,f\right)
\frac{L_3^2}{\sin^2\tht}+\\[5mm]\dst \hspace{8cm}
+\alf\,\frac{\sin\tht\,\sqrt{f}}{(\cos^2\tht)^{1/3}}\,\cos\phi
+\frac{\be}{(\cos^2\tht)^{1/3}},\\[5mm]\dst 
Q=-\frac 49\,L_3^3+2H\,L_3+3\alf\,(\cos\tht)^{1/3}\Big(\sqrt{f}\,L_1
+(\sqrt{f})'\,\cos\phi\,L_3\Big),\earr\right.\eeq
where $\,f(\tht)=\hat{f}(\cos\tht)$ with
\beq
\hat{f}(\mu)=\frac{\Big(\mu^{2/3}-\frac{\ze_1}{\ze_0}\Big)\Big(\mu^{2/3}-\frac{\ze_2}{\ze_0}\Big)}
{\mu^{4/3}+\mu^{2/3}+1},\qq \mu\in(-1,+1),\eeq
and $\,h(\tht)=\hat{h}(\cos\tht)$ with
\beq
\hat{h}(\mu)=-\mu^2+\frac 43\Big(1+\frac{\ze_1+\ze_2}{\ze_0}\Big)\mu^{4/3}
-2\Big(\frac{\ze_1+\ze_2}{\ze_0}+\frac{\ze_1\ze_2}{\ze_0^2}\Big)\mu^{2/3}
+4\frac{\ze_1\ze_2}{\ze_0^2}.\eeq
The parameter $\ze_0$ is:
\beq\label{ze0cas1}
\ze_0=\frac{\ze_1\ze_2}{(\sqrt{\ze_1}+\sqrt{\ze_2})^2}<\ze_1.\eeq
\erm
\end{nth}

\vspace{2mm}
\nin {\bf Proof of (i):}
The change of variable indicated gives (\ref{sys1qm}) by lengthy but straightforward computations.
$\quad\Box$

\nin{\bf Remark:} The previous analysis does not describe appropriately the special case 
$\ze_0=0$ for which elliptic functions are no longer required. In this case the coordinates change
\[\ze=\frac{\ze_1+\ze_2}{2}-\frac{\ze_1-\ze_2}{2}\,\cos\tht,\qq (\ze_1,\ze_2)\ \to\ (\pi,0),\]
gives for the metric
\beq
g=d\tht^2+\frac{\sin^2\tht}{1+\sin^2\tht\,G(\cos\tht)}\,d\phi^2,\eeq
with
\beq
G(\mu)=\frac{3\mu^2+4\rho\mu+1}{4(\rho+\mu)^2},\qq\rho=\frac{\ze_2+\ze_1}{\ze_2-\ze_1}>1.\eeq
The integrable system is
\beq\label{DM}\left\{\barr{l}\dst 
H=\frac 12\left(P_{\tht}^2+\Big(\frac 1{\sin^2\tht}+G(\cos\tht)\Big)\pf^2\right)
+\alf\,\frac{\sin\tht}{\sqrt{U}}\,\cos\phi+\frac{\be}{U},\\[5mm]\dst 
Q=-\pf^3+2H\,\pf+2\alf\,\sqrt{U}\sin\phi\,P_{\tht}
+2\alf\frac{(\sin\tht\sqrt{U})'}{\sin\tht}\,\cos\phi\,\pf.,\\[5mm]
U=\rho+\cos\tht,\earr\right.\eeq
on which we recognize the Dullin-Matveev system \cite{dm}.

\vspace{2mm}
\nin {\bf Proof of (ii):}
One has
\[G(0)=(\ze_1-\ze_2)^2\,\ze_0^2-2\ze_1\ze_2(\ze_1+\ze_2)\,\ze_0+\ze_1^2\ze_2^2.\]
Its vanishing determines uniquely $\ze_0$ in terms of $(\ze_1,\ze_2)$ as given by (\ref{ze0cas1}). 
At this stage positivity requires $\ze\in(0,\ze_0)$. Let us make the change of variable 
$\,\ze=\ze_0\,\mu^{2/3}$. The metric becomes
\[g=\frac 49\left\{\frac{d\mu^2}{(1-\mu^2)\hat{f}(\mu)}
+3(1-\mu^2)\frac{\hat{f}(\mu)}{\hat{h}(\mu)}\,d\phi^2\right\},\qq \mu\in(0,1).\]
All the functions in the metric are {\em even} functions of $\mu$: we can therefore 
take $\mu\in(-1,+1)$ extending the metric beyond $\mu=0$. One can check that the 
points $\mu=\pm 1$ are poles and therefore we get again for manifold ${\mb S}^2$. The change of 
variable $\mu=\cos\tht$ with $\tht\in(0,\pi)$ gives then for result (\ref{sys2qm}). $\quad\Box$

Let us proceed to:
\vspace{2mm}
\begin{nth} 
\brm
\item[(a)] The metric (\ref{metq}) for $\De_-=0$ is defined on ${\mb S}^2$ iff: 
\brm
\item[(i)] $\mbox{either}\quad F=\ze^2(\ze_0-\ze),\qq 0<\ze<\ze_0,\ $ and we have
\beq\label{sys2bisq}\left\{\barr{l}\dst 
H=\frac 12\Big(L_1^2+L_2^2+4L_3^2\Big)+\alf\,\sin\tht\,\cos\phi+\frac{\be}{\cos^2\tht},
\qq\tht\in(0,\pi),\\[5mm]\dst 
Q=-4\,L_3^3+2H\,L_3+\alf\Big(\cos\tht\,L_1-2\sin\tht\,\cos\phi\,L_3\Big),\earr\right.\eeq
which is the Goryachev-Chaplygin top.
\item[(ii)] $\mbox{or}\quad F=(\ze-\ze_1)^2(\ze_0-\ze)\ \mbox{and}\ \ G(0)=0,
\qq 0<\ze<\ze_0.$

\nin The integrable system is of the form (\ref{sys2qm}) with the functions 
\beq
\hat{f}(\mu)=\frac{\Big(4-\mu^{2/3}\Big)^2}
{\mu^{4/3}+\mu^{2/3}+1},\qq \hat{h}(\mu)=(4-\mu^{2/3})^3,\qq\mu\,\in(-1,+1).\eeq
\erm
\item[(b)]The metric (\ref{metq}) for $\De_-=0$ is defined on ${\mb H}^2$ iff:
\[F=(\ze-\ze_1)^2(\ze_0-\ze),\qq 0<\ze_1<\ze<\ze_0.\]
The integrable system, in the notations of Theorem 2, case (ii), is
\beq\label{sys2terq}\left\{\barr{l}\dst 
H=\frac 1{2\ze_-(u)}\left\{M_1^2+M_2^2-\left(1-\frac 3{C^2}\right)M_3^2\right\}+\alf\,\frac{T(1-T^2)}{\ze_-(u)}\,\cos\phi
+\frac{\be}{\ze_-(u)},\\[5mm]\dst
Q=-4M_3^3+2H\,M_3+\alf\Big(M_1-3T\,\cos\phi\,M_3\Big),\\[5mm]
\ze_-(u)=\rho-\tanh^2 u,\quad u\in(0,+\nf),\qq \rho=\frac{\ze_0}{\ze_0-\ze_1}>1.
\earr\right.\eeq
\erm
\end{nth}

\vspace{2mm}
\nin {\bf Proof of (a)(i):}
We have $F=\ze^2(\ze_0-\ze)$ and $G=\ze^3(4\ze_0-3\ze)$ and $\ze\in(0,\ze_0)$ from positivity. 
Taking for new variable $\tht$ such that $\ze=\ze_0\,\cos^2\tht$ we get 
for the metric
\beq
g=4\left(d\tht^2+\frac{\sin^2\tht}{1+3\sin^2\tht}\,d\phi^2\right),\qq \tht\in(0,\pi/2).\eeq
As it stands the manifold is $P^2({\mb R})$ (see \cite{Be}[p. 268]). However we can also 
extend the metric taking $\tht\in(0,\pi)$: then the manifold extends to ${\mb S}^2$ 
since $\tht=0$ and $\tht=\pi$ are poles and in this case we recover Goryachev-Chaplygin top. 
The observables can be transformed into (\ref{sys2bisq}). 
$\quad\Box$

\vspace{2mm}
\nin {\bf Proof of (a)(ii):}
In this case we have $\,G(0)=\ze_1^3(\ze_1-4\ze_0)$ which fixes $\ze_0=\ze_1/4$. The argument then proceeds as in the proof of Theorem 6, case (ii).$\quad\Box$

\vspace{2mm}
\nin {\bf Proof of (b):}
The proof is identical to the one for Theorem 5, case (ii), except for the change of coordinates, 
which is now
\[\ze=\ze_0-(\ze_0-\ze_1)\tanh^2 u:\qq (\ze_0,\ze_1)\quad\to\quad(0,+\nf).\]
One gets (\ref{sys2terq}) by similar arguments.$\quad\Box$

\begin{nth}
The metric (\ref{metq}) for $\De_->0$ is defined on ${\mb S}^2$ iff: 
\[ F=(\ze_0-\ze)(\ze-\ze_1)(\ze-\overline{\ze_1})\ \ \mbox{and}\ \  G(0)=0,
\quad 0<\ze<\ze_0.\]
The integrable system is of the form (\ref{sys2qm}) with the functions
\beq
\hat{f}(\mu)=\frac{\Big(\mu^{2/3}-\frac{\ze_1}{\ze_0}\Big)\Big(\mu^{2/3}-\frac{\ol{\ze}_1}{\ze_0}\Big)}{\mu^{4/3}+\mu^{2/3}+1},\qq\mu\,\in(-1,+1),\eeq
and
\beq
\hat{h}(\mu)=-\mu^2+\frac 43\Big(1+\frac{\ze_1+\ol{\ze}_1}{\ze_0}\Big)\mu^{4/3}
-2\Big(\frac{\ze_1+\ol{\ze}_1}{\ze_0}+\frac{|\ze_1|^2}{\ze_0^2}\Big)\mu^{2/3}
+4\frac{|\ze_1|^2}{\ze_0^2}.\eeq
We have two possible values for $\ze_0$ which are
\beq\label{ze0cas3}
\ze_0=\frac{|\ze_1|^2}{\ze_1+\ol{\ze}_1 \pm 2|\ze_1|}.\eeq
\end{nth}

\vspace{2mm}
\nin{\bf Proof:} 
We have
\[G(0)=(\ze_1-\ol{\ze}_1)^2\,\ze_0^2-2(\ze_1+\ol{\ze}_1)|\ze_1|^2\,\ze_0+|\ze_1|^4.\] 
Its vanishing gives for $\ze_0$ the roots (\ref{ze0cas3}). The subsequent analysis is identical 
to that already given in the proof of Theorem 6, case (ii).$\quad\Box$

It is interesting to examine the explicitly known integrable systems, with a metric 
defined in ${\mb S}^2$ and with a cubic observable already given in the literature:
\brm
\item The Goryachev-Chaplygin top given by Theorem 6, case (ii). 
\item The Dullin-Matveev top \cite{dm} is given in the remark after Theorem 6. 
\item  If we restrict, in Theorem 8, the parameters according to
\[\ze_0=-(\ze_1+\ol{\ze}_1)\quad\mbox{and}\quad \ze_0^2=|\ze_1|^2,
\qq\Longrightarrow\qq f=1,\quad g=4-\mu^3,\] 
we recover the Goryachev top
\beq\left\{\barr{l}\dst 
H=\frac 12\left(L_1^2+L_2^2+\frac 43\,L_3^2\right)+\alf\,\frac{\sin\tht}{(\cos^2\tht)^{1/3}}\,\cos\phi
+\frac{\be}{(\cos^2\tht)^{1/3}},\\[5mm]\dst
Q=-\frac 49\,\pf^3+2H\,\pf+3\alf\,(\cos\tht)^{1/3}\,L_1.\earr\right.\eeq 
\erm
The two new examples given by Tsiganov in \cite{Ts} are not defined on a manifold. 

\nin{\bf Remarks:}
\brm
\item All of the previous examples  belong to the third set with $p<0$.
\item Considering the genus of the algebraic curve 
$\dst y^2=\frac{F(\ze)}{\ze}$ let us observe that the Goryachev-Chaplygin and Dullin-Matveev 
systems have zero genus while the Goryachev system has genus one.
\item In general the potential $V$ as well as the observable $Q$ are not defined 
on the whole manifold. 
\erm

\section{Conclusion}
We have exhaustively constructed all the integrable models, on two dimensional manifolds, 
characterized by the following form of the observables
\beq\left\{\barr{l}\dst 
H=\frac 12\Big(\pth^2+a(\tht)\pf^2\Big)+f(\tht)\,\cos\phi+g(\tht)\\[4mm]\dst
Q=p\,\pf^3+q\,\Big(\pth^2+a(\tht)\pf^2\Big)\pf+\chi(\tht)\sin\phi\,\pth+\Big(\be(\tht)
+\ga(\tht)\cos\phi\Big)\pf\earr\right.\eeq
The main lesson from the failure of \cite{Se} to solve the problem has to do with the 
crucial role of the coordinates choice, which determines the structure of the ODE to be solved 
eventually. This is a familiar phenomenon to people dealing with Einstein equations: despite their  diffeomorphism invariance, finding exact solutions relies on an adapted choice of 
coordinates which can simplify, or even linearize the differential system to be integrated. 

\vspace{2mm}
\nin{\bf Acknowledgments:} we are greatly indebted to K. P. Tod for his kind and efficient 
help in the analysis of the metrics singularities of Section 5.

\end{document}